\documentclass[runningheads]{llncs}
\usepackage[T1]{fontenc}
\usepackage{todonotes}
\usepackage{hyperref}
\usepackage{hyperxmp}
\usepackage{graphicx}
\usepackage{url}
\usepackage{multirow}
\usepackage{subcaption}
\usepackage{booktabs} 
\usepackage{xspace} 
\usepackage{xcolor} 
\usepackage{colortbl} 
\usepackage{amsmath}

\newcommand{\modelname}{LLMDiRec\xspace}

\begin{document}
\title{LLMDiRec: LLM-Enhanced Intent Diffusion for Sequential Recommendation}
%

\author{Bo-Chian Chen\inst{1} \and
Manel Slokom\inst{2}\orcidID{0000-0002-9048-1906}
}

\authorrunning{Bo-Chian Chen and Manel Slokom}
\institute{Vrije Universiteit Amsterdam 
\email{jack.chen4@student.uva.nl}
\and
Centrum Wiskunde \& Informatica, Amsterdam,
\email{manel.slokom@cwi.nl}}
\maketitle              
\begin{abstract}
Existing sequential recommendation models, even advanced diffusion-based approaches, often struggle to capture the rich semantic intent underlying user behavior, especially for new users or long-tail items. This limitation stems from their reliance on ID-based embeddings, which lack semantic grounding. We introduce \modelname
, a new approach that addresses this gap by integrating Large Language Models (LLMs) into an intent-aware diffusion model. Our approach combines collaborative signals from ID embeddings with rich semantic representations from LLMs, using a dynamic fusion mechanism and a multi-task objective to align both views. We run extensive experiments on five public datasets. 
We run extensive experiments on five public datasets. 
We demonstrate that \modelname outperforms state-of-the-art algorithms, with particularly strong improvements in capturing complex user intents and enhancing recommendation performance for long-tail items.
\end{abstract}

\keywords{Sequential Recommendation, Diffusion Models, Large Language Models, Contrastive Learning, User Intent}

\section{Introduction}

Conventional recommender systems, such as collaborative filtering, typically model user preferences from static interactions, largely ignoring temporal dynamics. In contrast, sequential recommendation (SR) systems leverage the chronological order of user-item interactions to capture evolving preferences and predict the next item a user is likely to engage with \cite{wang2019survey,pan2025survey,fang2020sequential}. The field has evolved from early Markov chain models to deep learning approaches, including RNN-based models \cite{gru4rec} and Transformer-based architectures \cite{sasrec,bert4rec}.

Recent efforts focus on understanding the motivations behind user behavior. Intent-aware sequential recommenders~\cite{Qu2025,qin2024icsrec,chen2024intentenhanceddataaugmentationsequential,huang2024multiintentawarecontrastivelearning,wang2023intent} model latent user intents via contrastive learning or diffusion processes and achieve strong performance. However, these methods rely heavily on ID-based embeddings, which struggle to capture semantic intent, particularly for cold-start users and long-tail items. 
For instance, InDiRec clusters sequences into intent prototypes using K-means but lacks semantic grounding, risking misalignment between inferred intents and actual user motivations (e.g., purchasing a ``gift'' versus ``personal use'').
This limitation becomes especially problematic when collaborative signals are weak or user behavior is diverse. A user might interact with items such as ``keyboard,'' ``mouse,'' and ``monitor'' under a unified intent of ``assembling a computer.'' Random data augmentation in contrastive learning can disrupt this latent intent structure \cite{cadirec,Qu2025}, producing noisy training signals.

\begin{figure}[!htb]
    \centering
    \includegraphics[width=0.8\linewidth]{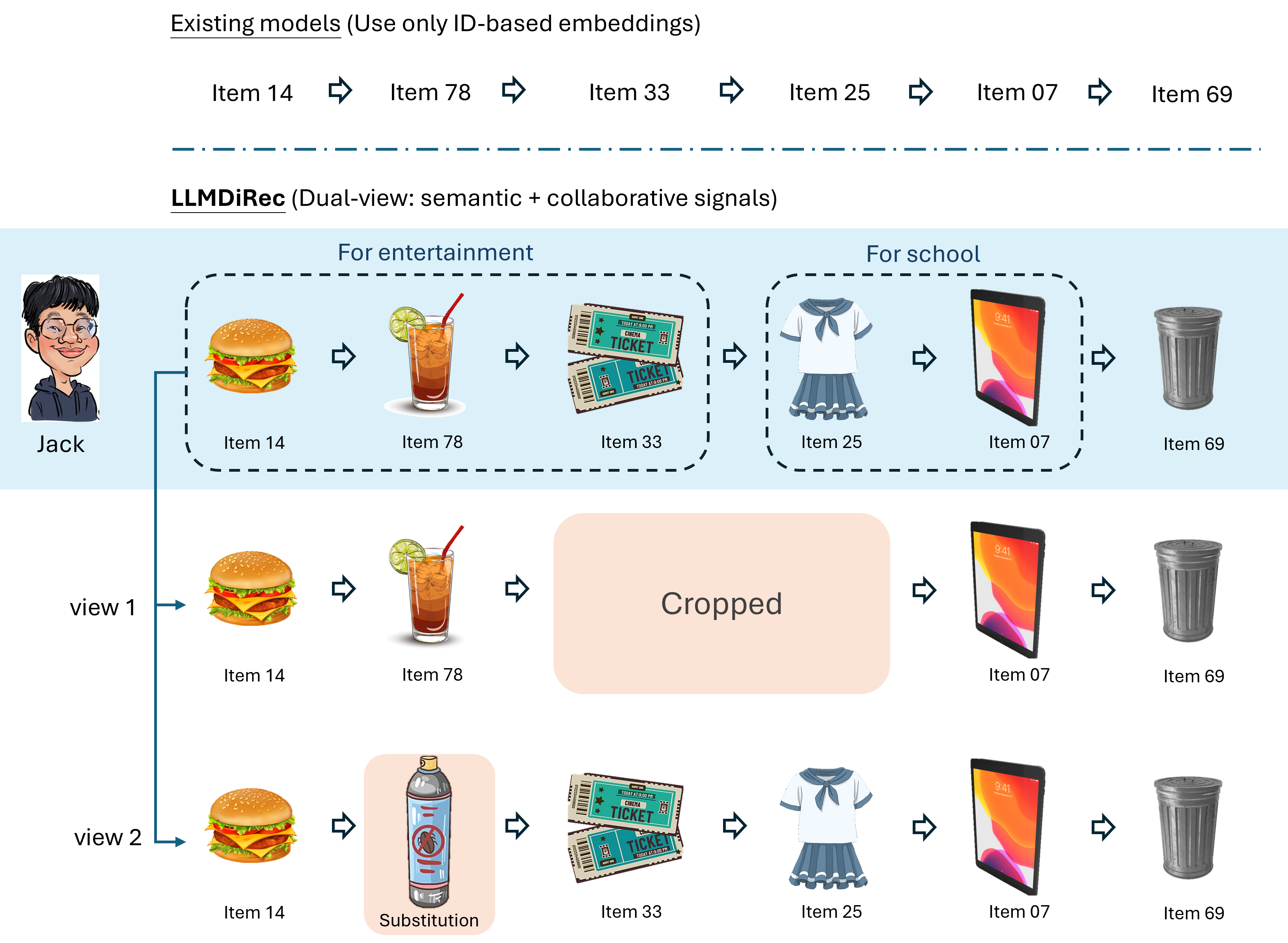}
    \caption{An illustration of semantically enhanced dual-view model and how random data augmentation (view 1, view 2) can disrupt semantic consistency. A user's interaction sequence contains two distinct intents: "For entertainment" and "For school." Augmentation methods like cropping or substitution can break the coherence of these intents, leading to flawed intent representations}
    \label{fig:augmentation_problem}
\end{figure}

ID-based embeddings often group items by co-occurrence rather than semantic meaning, mixing unrelated intents and missing the true structure. For example, a sequence containing a "gaming mouse" and a "textbook" might cluster with electronics due to the mouse, while the textbook relates to school supplies. Similarly, items like a uniform and an iPad may appear unrelated collaboratively but share a semantic intent ``for school.'' Without semantic grounding, models like InDiRec may fail to reflect real user motivations (Figure~\ref{fig:augmentation_problem}).
Semantic blindness leads to several challenges. ID-based models misalign intents by relying on co-occurrence rather than meaning. Cold-start users or items provide insufficient interaction data for reliable inference, leaving new items undiscovered. Long-tail items are disadvantaged, reinforcing popularity bias and reducing personalization \cite{Zhou_2023,WEI201729,meng2019wassersteincollaborativefilteringitem,PAN20202019EDP7258,10.1145/2964797.2964819}. Furthermore, random data augmentation in contrastive learning can destroy semantic coherence, mixing distinct intents like ``for entertainment'' and ``for school,'' resulting in flawed representations.

To address these challenges, we propose \textbf{LLMDiRec}, an enhanced version of InDiRec that integrates semantic embeddings from large language models (LLMs) into an intent-aware diffusion framework. Each item is represented by both collaborative ID embeddings and LLM-based semantic embeddings, enabling richer preference modeling even under sparsity. LLMDiRec employs dynamic fusion strategies, gated fusion, weighted combination, and cross-attention, to adaptively merge collaborative and semantic signals. To preserve intent coherence, user sequences are segmented into prefix-like subsequences, encoded, and clustered into prototypes, supporting effective aggregation across multi-intent behaviors. The model jointly optimizes next-item prediction, diffusion denoising, contrastive learning, and embedding alignment.
Our contributions are fourfold: (1) We propose the first framework to integrate LLM-based semantics into intent-aware diffusion for sequential recommendation; (2) We introduce a dual-view modeling approach with dynamic fusion that adapts to varying sparsity; (3) We achieve consistent state-of-the-art results on five datasets; and (4) We show large gains for long-tail items (up to 113\%) and provide insights into the role of semantics in intent modeling, validating LLMDiRec for practical use.

\section{Related Work}
\paragraph{Sequential Recommendation with Diffusion Models}
Standard sequential recommendation models, such as the self-attention based SASRec \cite{Kang2018}, have set strong baselines by capturing sequential dependencies. Other approaches have leveraged contrastive learning to enhance representation quality and address data sparsity issues \cite{Qiu2022,icl}. Recent works have explored diffusion models \cite{sohl2015deep,ho2020denoising} for recommendation, leveraging their generative capabilities to create high-quality training samples. CaDiRec \cite{Cui2024} applies diffusion to generate context-aware augmented views, while InDiRec \cite{Qu2025} introduces intent-aware diffusion with K-means clustering to guide the denoising process. However, these methods operate solely on ID-based embeddings, making the generative process semantically blind. Other approaches like DiffuASR \cite{diffuasr} and DreamRec \cite{dreamrec} also explore diffusion for data augmentation or direct generation, highlighting the growing interest in this area. DiffuRec \cite{li2023diffurec} demonstrates the potential of diffusion for collaborative filtering but lacks sequential modeling capabilities. \modelname extends this line of work by conditioning the diffusion process on semantically-rich signals derived from LLMs, leading to more meaningful and intent-consistent augmentations.

\paragraph{LLM Integration in Recommendation}
Large Language Models are increasingly being integrated into recommendation systems through various approaches \cite{lin2024recommendersystemsbenefit,wu2024survey}. Early works focus on using LLMs as feature extractors \cite{Liu2024llm-esr} or for generating item descriptions. 
LLM-ESR~\cite{Liu2024llm-esr} tackles the long-tail problem by initializing item features with LLM embeddings, showing strong performance on sparse items. More recent work explores semantic–collaborative alignment~\cite{Ren2024}, but typically relies on static objectives. In contrast, \modelname\ integrates LLM-based semantics directly into the generative diffusion process, enabling dynamic, intent-aware guidance throughout training, beyond treating LLM knowledge as auxiliary input.

\paragraph{Intent Modeling in Sequential Recommendation}
In sequential recommendation, user intent refers to the underlying purpose or goal behind a user's sequence of interactions \cite{icl}. Modeling this intent has been a central challenge. Traditional approaches like MIND \cite{li2019multiinterestnetworkdynamicrouting} use attention mechanisms to capture multiple user interests. Intent Contrastive Learning (ICL) \cite{icl} introduces clustering-based intent discovery with contrastive objectives, and InDiRec \cite{Qu2025} further advances this by using diffusion models to generate intent-aware augmentations. Recent works like ComiRec \cite{cen2020controllablemultiinterestframeworkrecommendation}, disentangled graph models \cite{wang2020disentangled}, and intent-interest disentanglement approaches \cite{choi2025idcl} explore controllable multi-interest frameworks. However, these methods rely primarily on collaborative signals for intent discovery. Our work combines semantic knowledge from LLMs with diffusion-based intent modeling to achieve more interpretable and semantically coherent representations.

\section{Our LLMDiRec Recommender}
\begin{figure}[]
    \centering
    \includegraphics[width=0.9\columnwidth]{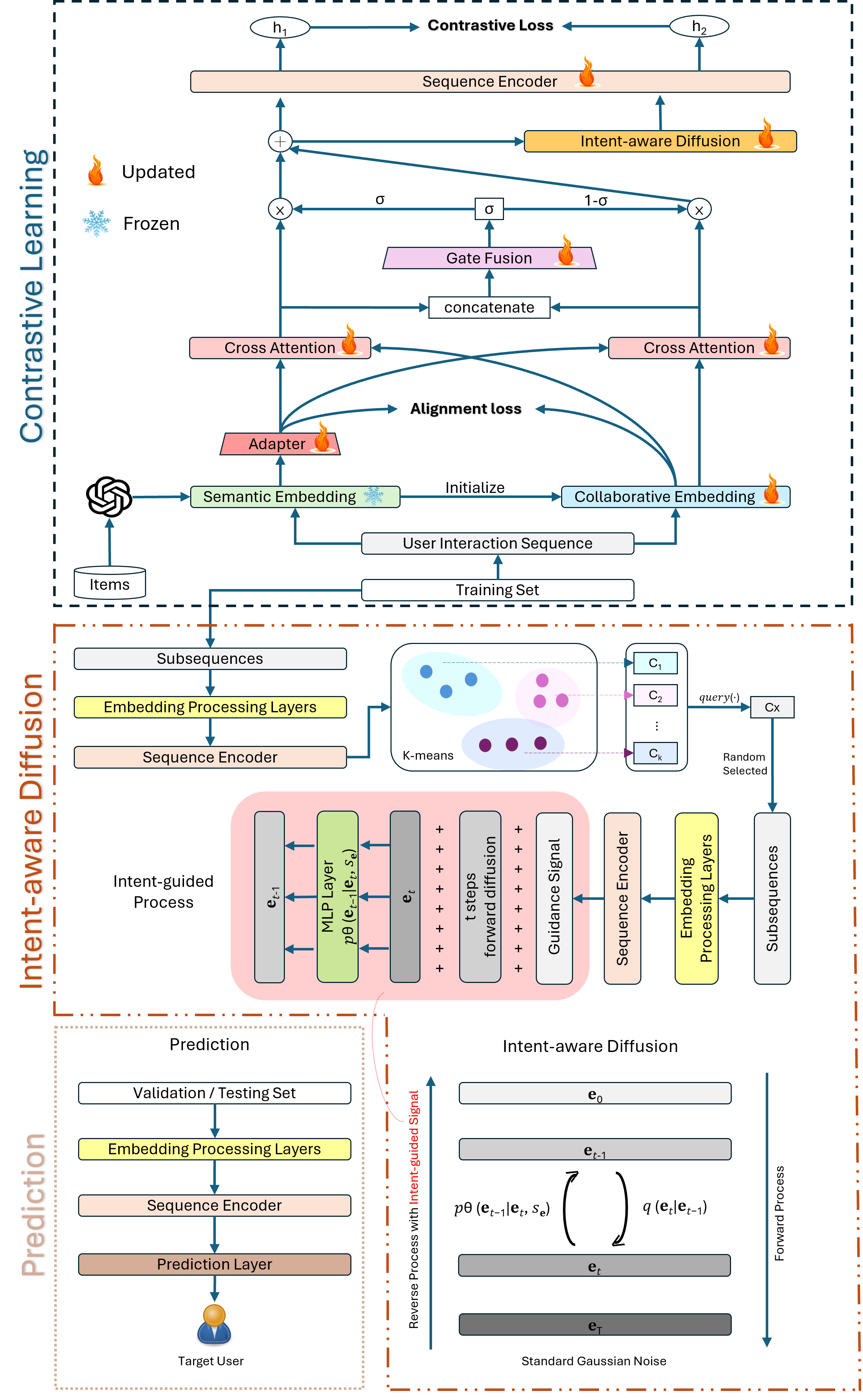} 
    \caption{Architecture of \modelname. The model fuses collaborative and semantic embeddings, encodes the sequence, and uses intent clusters to guide a diffusion process for contrastive learning, all optimized under a multi-task objective.}
    \label{fig:architecture}
\end{figure}

We provide a detailed overview of LLMDiRec in Figure~\ref{fig:architecture}. LLMDiRec comprises three main components (or phases): Dual-View Item Representation, Intent-Aware Diffusion with Semantic Guidance, and Multi-Task Optimization. 
\subsection{Dual-View Item Representation}
We model user preferences from two complementary views to capture both collaborative patterns and semantic relationships. Each item $i$ is represented by:  
(1) \textbf{Collaborative Embedding} $\mathbf{e}_i^{\text{id}} \in \mathcal{R}^d$, a standard trainable vector that captures interaction patterns and collaborative filtering signals from user-item data.  
(2) \textbf{Semantic Embedding} $\mathbf{e}_i^{\text{llm}} \in \mathcal{R}^{d'}$, a rich representation obtained by feeding carefully designed item descriptions into a pre-trained LLM (BAAI/bge-m3\footnote{\url{https://huggingface.co/BAAI/bge-m3}}\cite{Chen2024bge}). This embedding encapsulates semantic knowledge about item attributes, categories, and contextual relationships~\cite{Liu2024llm-esr}.


To bridge the semantic-collaborative gap, we employ a lightweight adapter network that projects $\mathbf{e}_i^{\text{llm}}$ into the model's embedding space while preserving semantic structure. The semantic embedding is kept frozen during training to maintain its rich pre-trained knowledge. We then combine these dual views using a learnable gated fusion mechanism:
$\mathbf{e}_i = \boldsymbol{\gamma} \odot \mathbf{e}_i^{\text{id}} + (1-\boldsymbol{\gamma}) \odot \text{Adapter}(\mathbf{e}_i^{\text{llm}})$, where $\boldsymbol{\gamma} = \sigma(\mathbf{W}[\mathbf{e}_i^{\text{id}}; \text{Adapter}(\mathbf{e}_i^{\text{llm}})])$ is a learned gating vector that adaptively balances the two views based on item characteristics.

\subsection{Intent-Aware Diffusion with Semantic Guidance}
Building upon InDiRec \cite{Qu2025}, we leverage a conditional diffusion process to generate high-quality positive samples for contrastive learning, enhanced with semantic guidance. The process consists of three key components:

\paragraph{Semantic-Enhanced Intent Discovery} We segment user sequences into subsequences using dynamic incremental prefix segmentation to capture fine-grained behavioral patterns. Each subsequence is encoded using our dual-view representations to obtain $\{\mathbf{h}_i\}$. We then apply K-means clustering on these semantically-enhanced representations to identify $K$ intent prototypes $\{\mathbf{c}_k\}$. Unlike purely collaborative approaches, our intent prototypes capture both interaction patterns and semantic relationships, resulting in an interpretable and coherent user motivations (e.g., "shopping for outdoor activities," "buying kitchen essentials").

\paragraph{Conditional Denoising with Intent Signals} We train a denoising network $f_\theta$ to reverse a fixed noising process, where the forward process gradually corrupts clean sequence representations $\mathbf{x}_0$ with Gaussian noise \cite{sohl2015deep,ho2020denoising}: $\mathbf{x}_t = \sqrt{\bar{\alpha}_t}\mathbf{x}_0 + \sqrt{1-\bar{\alpha}_t}\boldsymbol{\epsilon}$. The denoising is conditioned on intent signal $\mathbf{s}$, which corresponds to the nearest prototype $\mathbf{c}_k$ for the input sequence. The diffusion loss is:
$\mathcal{L}_{\text{diff}} = \mathrm{E}_{\mathbf{x}_0, t, \boldsymbol{\epsilon}} [\|\boldsymbol{\epsilon} - f_\theta(\mathbf{x}_t, \mathbf{s}, t)\|^2]$.

\paragraph{Intent-Consistent Augmentation} During inference, we sample intent-consistent augmentations by conditioning the denoising process on specific intent prototypes, ensuring that generated sequences maintain semantic coherence with the original user behavior patterns.
To sum up this phase, the semantics enhanced diffusion process produces augmentations that are not only diverse but also semantically meaningful and intent-consistent, thereby improving the quality of contrastive learning.

\subsection{Multi-task Optimization}
\modelname is trained end-to-end by jointly optimizing a multi-task objective. As depicted in Figure~\ref{fig:architecture}, the total loss $\mathcal{L}$ is a weighted sum of four components:
\begin{equation}
\mathcal{L} = \lambda_{\text{rec}}\mathcal{L}_{\text{rec}} + \lambda_{\text{diff}}\mathcal{L}_{\text{diff}} + \lambda_{\text{cl}}\mathcal{L}_{\text{cl}} + \lambda_{\text{align}}\mathcal{L}_{\text{align}}
\label{eq:total_loss}
\end{equation}
where:
\begin{itemize}
        \item $\mathcal{L}_{\text{rec}}$ is the standard cross-entropy loss for next-item prediction.
        \item $\mathcal{L}_{\text{diff}}$ is the diffusion denoising loss described above.
        \item $\mathcal{L}_{\text{cl}}$ is an InfoNCE contrastive loss \cite{Ren2024} that pulls representations of an original sequence and its diffused augmentation closer.
        \item $\mathcal{L}_{\text{align}}$ is a cosine similarity loss that enforces alignment between the collaborative embedding $\mathbf{e}_i^{\text{id}}$ and the semantic embedding $\mathbf{e}_i^{\text{llm}}$ of the same item, facilitating knowledge transfer.
\end{itemize}

\section{Experimental Setup}
\label{sec:exp_setup}
To evaluate the effectiveness of our proposed LLM-enhanced InDiRec framework, we conduct experiments on multiple public sequential recommendation datasets. We compare our method with a range of state-of-the-art baselines.

\subsection{Datasets}
\label{sec:datasets}
We evaluate our method on five widely-used public datasets from different domains. Three datasets (Beauty, Sports, and Toys) originate from Amazon\footnote{\url{https://jmcauley.ucsd.edu/data/amazon/}}'s e-commerce platform, representing different product categories with rich user-item interaction histories. The Yelp\footnote{\url{https://www.yelp.com/dataset}} dataset captures user preferences in local business recommendations, while MovieLens-1M\footnote{\url{https://grouplens.org/datasets/movielens/}} (ML-1M) provides movie rating data and serves as a standard benchmark in recommendation research. 
Following established preprocessing protocols~\cite{Liu2024llm-esr}, we filter users and items with fewer than five interactions to ensure data quality. The resulting datasets are highly sparse, which is a common characteristic of real-world recommendation scenarios. 
Detailed statistics for all datasets are presented in Table~\ref{tab:dataset_stats}.
\begin{table}[!ht]
\centering
\begin{tabular}{l|ccccc}
\toprule
Datasets & Beauty & Sports & Toys & Yelp & ML-1M \\
\midrule
\# Users        & 22,363  & 35,598  & 19,412  & 15,720& 6,041   \\
\# Items        & 12,101  & 18,357  & 11,924  & 11,383& 3,417   \\
\# Avg. Length  & 8.9     & 8.3     & 8.6     & 12.23& 165.5   \\
\# Actions      & 198,502 & 296,337 & 167,597 & 192,214 & 999,611 \\
Sparsity        & 99.93\% & 99.95\% & 99.93\% & 99.89\%& 95.16\% \\
\bottomrule
\end{tabular}
\caption{Statistics of the experimental datasets.}
\label{tab:dataset_stats}
\end{table}
For training and evaluation, we use the standard leave-one-out splitting strategy, where the most recent interaction serves as the test set, the second most recent as validation, and the remainder for training.

\subsection{LLM Prompt Design}
\label{sec:prompt_design}

To generate high-quality semantic representations, we design prompts inspired by \cite{Liu2024llm-esr}, using a consistent structure tailored to each dataset's key attributes.
\begin{itemize}
\item \textbf{Amazon Beauty}: ``The beauty item has following attributes: \textbackslash n name is <TITLE>; brand is <BRAND>; price is <PRICE>. \textbackslash n The item has following features: <CATEGORIES>. \textbackslash n The item has following descriptions: <DESCRIPTION>.''

\item \textbf{Amazon Sports}: ``The Sports and Outdoors item has following attributes: \textbackslash n name is <TITLE>; brand is <BRAND>; price is <PRICE>. \textbackslash n The item has following features: <CATEGORIES>. \textbackslash n The item has following descriptions: <DESCRIPTION>.''

\item \textbf{Amazon Toys}: ``The Toys \& Games item has following attributes: \textbackslash n name is <TITLE>; brand is <BRAND>; price is <PRICE>. \textbackslash n The item has following features: <CATEGORIES>. \textbackslash n The item has following descriptions: <DESCRIPTION>.''

\item \textbf{Yelp}: ``The point of interest has the following attributes: \textbackslash n name is <NAME>; category is <CATEGORY>; type is <TYPE>; open status is <OPEN>; review count is <COUNT>; city is <CITY>; average score is <STARS>.''

\item \textbf{MovieLens-1M}: ``The movie item has following attributes: \textbackslash n Title: <TITLE> \textbackslash n Genres: <GENRES> \textbackslash n Year: <YEAR>''
\end{itemize}
These prompts are processed by a pre-trained language model (e.g., bge-m3, text-embedding-ada-002) to generate dense semantic embeddings that capture rich contextual information beyond traditional collaborative filtering signals.

\subsection{Baselines and Evaluation Metrics}
\label{sec:baselines}
We compare \textbf{LLMDiRec} with representative methods across key modeling paradigms:  
(1) \textit{General models}: \textbf{SASRec}~\cite{sasrec}, a foundational self-attention-based sequential recommender.  
(2) \textit{Contrastive learning models}: \textbf{DuoRec}~\cite{duorec}, a dual-encoder model with model-level augmentation and supervised sampling.  
(3) \textit{Diffusion-based models}: \textbf{CaDiRec}~\cite{cadirec}, which uses conditional diffusion for contrastive learning; and \textbf{InDiRec}~\cite{Qu2025}, which combines intent-aware diffusion and contrastive learning, most closely related to our approach.


We evaluate the models using standard top-$k$ recommendation metrics:  
\textbf{Hit Rate@k (HR@k)}: The fraction of times the ground-truth item appears in the top-$k$ predictions.  
\textbf{NDCG@k}: Normalized Discounted Cumulative Gain at $k$, which accounts for the position of the correct item in the ranking.

\subsection{Implementation Details}
\label{sec:exp_implementation}
Experiments run on an Intel i7-1195G7 CPU with NVIDIA A6000 GPUs, using Python 3.11.11 and PyTorch 2.5.1+cu124. Our code is available on GitHub.
We use the Adam optimizer with a learning rate of 0.001 and batch size $\{256, 512\}$. Diffusion steps $\{10, 50, 100, 200\}$, clustering intervals $\{32, 64, 128, 256, 512, 1024\}$ are tuned on the validation set. LLM embeddings are generated using a pre-trained model. All models are trained for 100 epochs with early stopping.

\section{Recommendation Performance}
Table~\ref{tab:main_results} presents the main results of our experiments, comparing LLMDiRec with several strong sequential recommendation baselines (SASRec, DuoRec, CaDiRec, and InDiRec) across five benchmark datasets. We report HR@5, NDCG@5, HR@10, and NDCG@10 for each method and dataset.
\begin{table}[!htb]
\scriptsize
    \resizebox{\columnwidth}{!}{%
    \begin{tabular}{llccccrr}
        \toprule
        Dataset & Metric & SASRec & DuoRec & CaDiRec & InDiRec & \textbf{\modelname} & Improv. \\
        \midrule
        \multirow{2}{*}{ML-1M} 
        & HR@10 & 0.1902 & 0.1996 & 0.4092 & \underline{0.6291} & \textbf{0.6417} & +2.0\% \\
        & ND@10 & 0.0906 & 0.1003 & 0.2253 & \underline{0.4188} & \textbf{0.4300} & +2.7\% \\
        \midrule
        \multirow{2}{*}{Beauty} 
        & HR@10 & 0.0628 & 0.0669 & 0.0699 & \underline{0.0939} & \textbf{0.0971} & +3.4\% \\
        & ND@10 & 0.0321 & 0.0336 & 0.0385 & \underline{0.0572} & \textbf{0.0589} & +3.0\% \\
        \midrule
        \multirow{2}{*}{Sports} 
        & HR@10 & 0.0339 & 0.0362 & 0.0404 & \underline{0.0550} & \textbf{0.0579} & +5.3\% \\
        & ND@10 & 0.0174 & 0.0189 & 0.0218 & \underline{0.0322} & \textbf{0.0345} & +7.1\% \\
        \midrule
        \multirow{2}{*}{Toys} 
        & HR@10 & 0.0675 & 0.0681 & 0.0766 & \underline{0.1004} & \textbf{0.1058} & +5.4\% \\
        & ND@10 & 0.0374 & 0.0385 & 0.0439 & \underline{0.0622} & \textbf{0.0662} & +6.4\% \\
        \midrule
        \multirow{2}{*}{Yelp} 
        & HR@10 & 0.0274 & 0.0342 & 0.0451 & \underline{0.0572} & \textbf{0.0607} & +6.1\% \\
        & ND@10 & 0.0136 & 0.0189 & 0.0232 & \underline{0.0300} & \textbf{0.0315} & +5.0\% \\
        \bottomrule
    \end{tabular}
    }
    \caption{Main performance comparison. Best results are in \textbf{bold}, second-best are \underline{underlined}. Improv. is the relative gain of \modelname over the second-best baseline. All improvements are statistically significant (p < 0.05).}
    \label{tab:main_results}
\end{table}


We observe that LLMDiRec model achieves the best performance across all datasets and metrics. 
The improvements are especially pronounced on sparser datasets such as Sports, Toys, and Yelp, where LLMDiRec outperforms the strongest baseline, InDiRec, by 5–8\% in HR@10 and NDCG@10. On the denser ML-1M dataset, LLMDiRec still provides consistent gains, with a 2–3\% relative improvement over InDiRec. These results indicate that integrating LLM-based semantic representations enhances recommendation performance, particularly in scenarios with limited collaborative signals.

The consistent outperformance across datasets and metrics demonstrates that LLMDiRec effectively leverages both collaborative signals and semantic knowledge, yielding more accurate and robust sequential recommendations. The larger gains on sparse datasets suggest that LLM embeddings are particularly beneficial when collaborative signals are weak, providing semantic understanding that supports better intent modeling and representation learning.

\section{Long-tail and Cold-start Performance Analysis}
Table~\ref{tab:longtail_coldstart} presents analysis of \modelname's performance across different user and item groups, demonstrating its effectiveness in addressing challenging scenarios where traditional collaborative methods struggle. 

\begin{table}[ht]
\centering
\resizebox{\columnwidth}{!}{
\begin{tabular}{ll|cc|cc|cc|cc|cc}
\toprule
\multirow{2}{*}{Dataset} & \multirow{2}{*}{Model} & \multicolumn{2}{c|}{Overall} & \multicolumn{2}{c|}{Tail Item} & \multicolumn{2}{c|}{Head Item} & \multicolumn{2}{c|}{Cold User} & \multicolumn{2}{c}{Hot User} \\
 & & HR@10 & N@10 & HR@10 & N@10 & HR@10 & N@10 & HR@10 & N@10 & HR@10 & N@10 \\
\midrule
\multirow{3}{*}{ML-1M}
& CaDiRec & 0.4092 & 0.2253 & \underline{0.1111} & \underline{0.0370} & 0.4170 & 0.2298 & 0.2968 & 0.1726 & 0.4667 & 0.2533 \\
& InDiRec & \underline{0.6291} & \underline{0.4188} & 0.0550 & 0.0322 & \underline{0.6291} & \underline{0.4188} & \textbf{0.6141} & \textbf{0.4101} & \underline{0.6075} & \underline{0.4072} \\
& \cellcolor{blue!10}LLMDiRec & \cellcolor{blue!10}\textbf{0.6313} & \cellcolor{blue!10}\textbf{0.4226} & \cellcolor{blue!10}\textbf{0.1429} & \cellcolor{blue!10}\textbf{0.0783} & \cellcolor{blue!10}\textbf{0.6445} & \cellcolor{blue!10}\textbf{0.4330} & \cellcolor{blue!10}\underline{0.4799} & \cellcolor{blue!10}\underline{0.3128} & \cellcolor{blue!10}\textbf{0.7109} & \cellcolor{blue!10}\textbf{0.4816} \\
\midrule
\multirow{3}{*}{Beauty}
& CaDiRec & 0.0699 & 0.0385 & 0.0190 & 0.0123 & 0.0711 & 0.0393 & 0.0631 & 0.0344 & 0.0786 & 0.0452 \\
& InDiRec & \underline{0.0939} & \underline{0.0572} & \underline{0.0218} & \underline{0.0124} & \underline{0.0992} & \underline{0.0596} & \textbf{0.0922} & \textbf{0.0557} & \underline{0.0966} & \underline{0.0579} \\
& \cellcolor{blue!10}LLMDiRec & \cellcolor{blue!10}\textbf{0.0971} & \cellcolor{blue!10}\textbf{0.0589} & \cellcolor{blue!10}\textbf{0.0291} & \cellcolor{blue!10}\textbf{0.0219} & \cellcolor{blue!10}\textbf{0.1027} & \cellcolor{blue!10}\textbf{0.0623} & \cellcolor{blue!10}\underline{0.0886} & \cellcolor{blue!10}\underline{0.0542} & \cellcolor{blue!10}\textbf{0.1204} & \cellcolor{blue!10}\textbf{0.0728} \\
\midrule
\multirow{3}{*}{Sports}
& CaDiRec & 0.0404 & 0.0218 & 0.0052 & 0.0034 & 0.0438 & 0.0235 & 0.0423 & 0.0229 & 0.0389 & 0.0206 \\
& InDiRec & \underline{0.0550} & \underline{0.0322} & \underline{0.0064} & \underline{0.0059} & \underline{0.0577} & \underline{0.0341} & \underline{0.0536} & \underline{0.0320} & \underline{0.0549} & \underline{0.0318} \\
& \cellcolor{blue!10}LLMDiRec & \cellcolor{blue!10}\textbf{0.0579} & \cellcolor{blue!10}\textbf{0.0345} & \cellcolor{blue!10}\textbf{0.0091} & \cellcolor{blue!10}\textbf{0.0069} & \cellcolor{blue!10}\textbf{0.0612} & \cellcolor{blue!10}\textbf{0.0364} & \cellcolor{blue!10}\textbf{0.0590} & \cellcolor{blue!10}\textbf{0.0355} & \cellcolor{blue!10}\textbf{0.0549} & \cellcolor{blue!10}\textbf{0.0322} \\
\midrule
\multirow{3}{*}{Toys}
& CaDiRec & 0.0766 & 0.0439 & 0.0279 & \underline{0.0154} & 0.0830 & 0.0479 & 0.0818 & 0.0474 & 0.0722 & 0.0413 \\
& InDiRec & \underline{0.1004} & \underline{0.0622} & \underline{0.0218} & 0.0124 & \underline{0.1073} & \underline{0.0669} & \underline{0.0993} & \underline{0.0605} & \textbf{0.1028} & \textbf{0.0661} \\
& \cellcolor{blue!10}LLMDiRec & \cellcolor{blue!10}\textbf{0.1058} & \cellcolor{blue!10}\textbf{0.0662} & \cellcolor{blue!10}\textbf{0.0464} & \cellcolor{blue!10}\textbf{0.0298} & \cellcolor{blue!10}\textbf{0.1115} & \cellcolor{blue!10}\textbf{0.0698} & \cellcolor{blue!10}\textbf{0.1091} & \cellcolor{blue!10}\textbf{0.0705} & \cellcolor{blue!10}\underline{0.1004} & \cellcolor{blue!10}\underline{0.0589} \\
\midrule
\multirow{3}{*}{Yelp}
& CaDiRec & 0.0451 & 0.0232 & 0.0009 & 0.0006 & 0.0582 & 0.0297 & 0.0434 & 0.0223 & 0.0451 & 0.0230 \\
& InDiRec & \underline{0.0572} & \underline{0.0300} & \underline{0.0028} & \underline{0.0014} & \underline{0.0747} & \underline{0.0388} & \underline{0.0575} & \underline{0.0297} & \textbf{0.0595} & \textbf{0.0315} \\
& \cellcolor{blue!10}LLMDiRec & \cellcolor{blue!10}\textbf{0.0607} & \cellcolor{blue!10}\textbf{0.0315} & \cellcolor{blue!10}\textbf{0.0048} & \cellcolor{blue!10}\textbf{0.0023} & \cellcolor{blue!10}\textbf{0.0764} & \cellcolor{blue!10}\textbf{0.0396} & \cellcolor{blue!10}\textbf{0.0603} & \cellcolor{blue!10}\textbf{0.0316} & \cellcolor{blue!10}\underline{0.0593} & \cellcolor{blue!10}\underline{0.0295} \\
\bottomrule
\end{tabular}
}
\caption{Long-tail and cold-start performance on five datasets. The best results are in bold, and the second-best are underlined. Our LLMDiRec performance is highlighted in light blue.}
\label{tab:longtail_coldstart}
\end{table}

\textbf{Long-tail Item Performance:} For tail items (bottom 20\% by popularity), \modelname achieves an improvements across all datasets. The most striking gains are observed on ML-1M with 160\% improvement (0.0550→0.1429 HR@10), Toys with 113\% improvement (0.0218→0.0464), and Beauty with 33\% improvement (0.0218→0.0291). Even on challenging datasets like Yelp, \modelname delivers 71\% improvement (0.0028→0.0048). These substantial gains confirm that semantic embeddings provide crucial signals where collaborative data is scarce, effectively addressing the long-tail item bias inherent in ID-based approaches.

\textbf{Cold-start User Analysis:} For cold-start users, \modelname demonstrates consistent improvements while maintaining competitive performance. On Toys, cold-start users achieve 0.1091 HR@10 with \modelname vs. 0.0993 with InDiRec, representing a 10\% improvement. On Sports, the improvement reaches 10\% (0.0536$\rightarrow$0.0590), while Beauty shows modest but consistent gains. Notably, \modelname maintains strong performance for hot users as well, indicating that semantic enhancement doesn't compromise collaborative signal utilization for users with rich interaction histories.

\textbf{Head Item and Hot User Performance:} \modelname also improves recommendations for popular items and active users, showing that semantic knowledge complements rather than replaces collaborative signals. For head items, improvements range from 4-8\% across datasets, while hot users see 2-17\% gains, demonstrating the universal applicability of our approach.

\section{Ablation Study and Component Analysis} 
To understand the contribution of each component in \textbf{LLMDiRec}, we conduct ablation studies by systematically removing or modifying key modules.
\begin{table}[!ht]
\centering
\resizebox{\textwidth}{!}{
\begin{tabular}{l|l|cccccc}
\toprule
 & Metric & w/ CA & w/ CA, w/o $\mathcal{L}_{align}$ & w/o $\mathcal{L}_{align}$ & w/ concat & w/ weighted & LLMDiRec \\
\midrule
\multirow{2}{*}{ML-1M}
& HR@10   & 0.6486 & 0.6478 & 0.6416 & 0.5883 & 0.6306 & 0.6417 \\
& ND@10   & 0.4369 & 0.4375 & 0.4259 & 0.3970 & 0.4277 & 0.4300 \\
\midrule
\multirow{2}{*}{Beauty}
& HR@10   & 0.0741 & 0.0716 & 0.0933 & 0.0885 & 0.0931 & 0.0971 \\
& ND@10   & 0.0411 & 0.0393 & 0.0567 & 0.0507 & 0.0566 & 0.0589 \\
\midrule
\multirow{2}{*}{Sports}
& HR@10   & 0.0287 & 0.0183 & 0.0554 & 0.0519 & 0.0536 & 0.0579 \\
& ND@10   & 0.0152 & 0.0090 & 0.0321 & 0.0288 & 0.0314 & 0.0345 \\
\midrule
\multirow{2}{*}{Toys}
& HR@10   & 0.0442 & 0.0413 & 0.1002 & 0.0966 & 0.1002 & 0.1058 \\
& ND@10   & 0.0239 & 0.0223 & 0.0625 & 0.0564 & 0.0631 & 0.0662 \\
\midrule
\multirow{2}{*}{Yelp}
& HR@10   & 0.0461 & 0.0473 & 0.0587 & 0.0620 & 0.0607 & 0.0607 \\
& ND@10   & 0.0232 & 0.0242 & 0.0307 & 0.0330 & 0.0318 & 0.0315 \\
\bottomrule
\end{tabular}
}
\caption{Ablation study of LLMDiRec on five datasets.}
\label{tab:llmdirec_ablation}
\end{table}
Table~\ref{tab:llmdirec_ablation} reports HR@10 and NDCG@10 across five datasets for the following variants:
\begin{itemize}
    \item \textbf{w/ CA}: Uses cross-attention fusion with all loss components.
    \item \textbf{w/ CA, w/o $\mathcal{L}_{align}$}: Cross-attention fusion without alignment loss.
    \item \textbf{w/o $\mathcal{L}_{align}$}: Baseline fusion without cross-attention or alignment loss.
    \item \textbf{w/ concat}: Uses simple concatenation-based fusion.
    \item \textbf{w/ weighted}: Applies weighted fusion of embeddings.
    \item \textbf{LLMDiRec}: Full model with all components.
\end{itemize}

\textbf{Cross-Attention Fusion:} The effectiveness of cross-attention (CA) is dataset-dependent. On ML-1M, it improves performance (HR@10: 0.6486 vs. 0.6416), while on others like Toys, it significantly degrades it (0.0442 vs. 0.1002, a 55.9\% drop). This reflects ML-1M's longer sequences (165.5 vs. 8–12), lower sparsity (95.16\% vs. 99.9+\%), and domain-specific temporal patterns, which favor attention mechanisms. In contrast, e-commerce datasets benefit from simpler fusion, which avoids overfitting under sparse, short histories.

\textbf{Alignment Loss:} Comparing ``w/o $\mathcal{L}_{align}$'' to the full LLMDiRec shows consistent gains on Beauty (+4.1\%), Sports (+4.5\%), and Toys (+5.6\%). While alignment can hurt performance when combined with CA, our model applies it effectively, improving results across most datasets. On Yelp, performance is unchanged, showing model adaptability.

\textbf{Fusion Strategy Comparison:} LLMDiRec consistently outperforms simpler strategies. Compared to concatenation, it improves results on Beauty (+9.7\%), Sports (+11.6\%), and Toys (+9.5\%), with similar results on Yelp. Compared to weighted fusion, gains remain consistent, highlighting LLMDiRec's strength in integrating collaborative and semantic signals.

\textbf{Component Contribution:} Removing components reveals their impact. The largest drops occur when replacing LLMDiRec with CA variants, especially on sparse datasets (e.g., over 100\% drop on Sports and Toys), showing the benefit of adaptive fusion. On ML-1M, differences are smaller (<3\%), reflecting robustness. Across all datasets, LLMDiRec achieves the best or near-best results (e.g., ML-1M: 0.6417 vs. 0.6486), confirming its generalizability across domains.

\section{Representation Quality and Interpretability}

\begin{figure}[]
    \centering
    \includegraphics[width=\columnwidth]{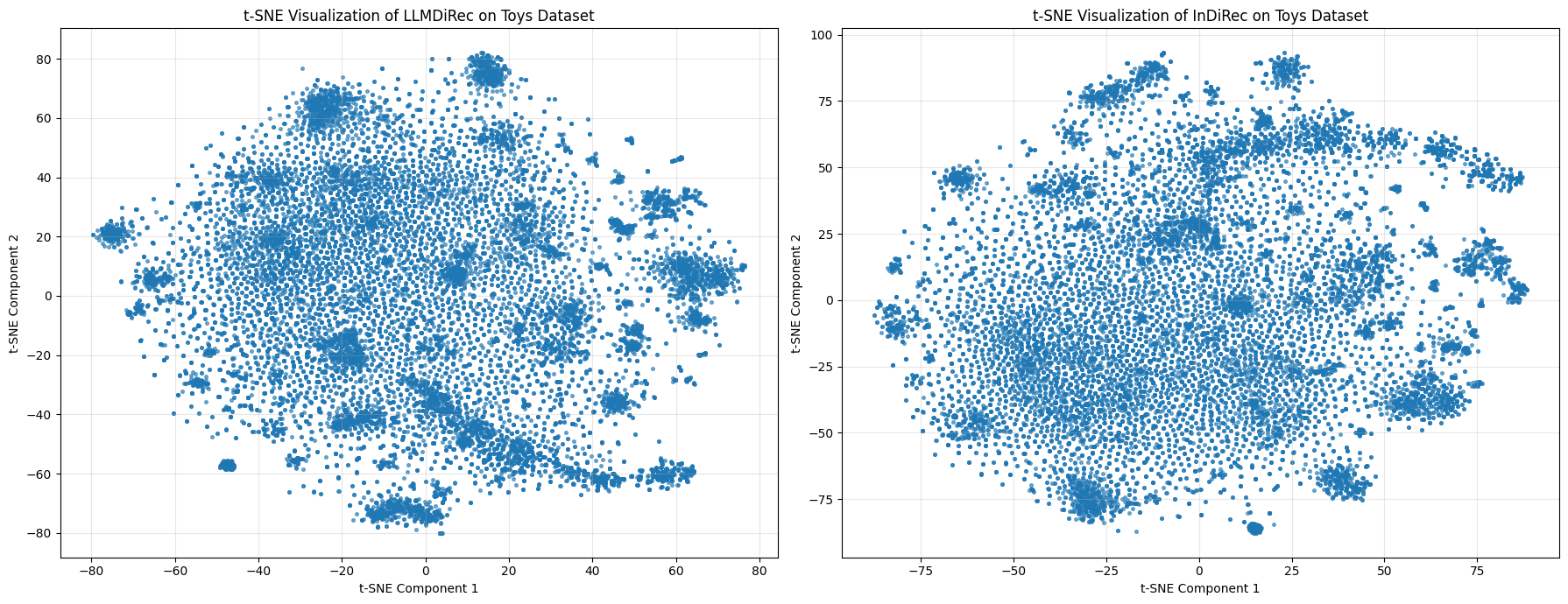}
    \caption{t-SNE visualization of learned sequence representations on the Toys dataset for InDiRec (left) and \modelname (right). LLM-enhanced representations form more distinct and coherent clusters.}
    \label{fig:tsne}
\end{figure}

Figure~\ref{fig:tsne} shows t-SNE visualizations comparing \modelname\ with InDiRec. \modelname\ learns more discriminative and semantically coherent representations, forming tighter, more distinct clusters.
This provides clearer separation of user intents and improved recommendation quality.
To quantify clustering quality, we compute silhouette scores: \modelname\ achieves 0.1197 on Toys vs. 0.0294 for InDiRec, a 307\% improvement, highlighting the impact of semantic enhancement on intent representation.

\section{Conclusion and Future Work}
We introduced \modelname, a new approach that integrates semantic knowledge from LLMs into an intent-aware diffusion model for sequential recommendation. By integrating collaborative and semantic signals, our model generates higher-quality augmentations for contrastive learning. Experiments on five datasets show that \modelname consistently achieves state-of-the-art performance, with significant gains for long-tail items (up to 113\%) and cold-start users (up to 10\%). Our analysis confirms that semantic enhancement leads to more coherent intent representations and robust performance, especially for space data with cold-start users and cold-start items.

Future work will explore more advanced fusion mechanisms, extend the framework to cross-domain recommendation, and develop tools for interpreting the learned intent prototypes. 
%
%
%
\bibliographystyle{splncs04}
\bibliography{paper}
\end{document}